\newcommand{\CLASSINPUTinnersidemargin}{17mm}
\documentclass[10pt,final,journal,letterpaper,oneside,twocolumn]{IEEEtran}
\usepackage{ucs}
\usepackage[english]{babel}
\usepackage[T1]{fontenc}
\usepackage{euscript}
\usepackage{amsmath}
\usepackage{amsfonts}
\usepackage{amssymb}
\usepackage{exscale}
\usepackage{graphics}
\usepackage{graphicx}
\usepackage{battail}
\usepackage{fancyhdr}
\usepackage{mathtext}
\usepackage{latexsym}
\usepackage{fancybox} %: Variants of \fbox
\usepackage{pstricks} %
\usepackage{lineno}
\usepackage{xcolor}
\usepackage{subcaption}
\usepackage{makecell}
\usepackage{wrapfig}

\author{Maxim Goryachev, Zeyu Kuang, Eugene N. Ivanov, Philipp Haslinger, Holger Muller, Michael E. Tobar
\thanks{Maxim Goryachev, Eugene N. Ivanov, Michael E. Tobar are with ARC Centre of Excellence for Engineered Quantum Systems, School of Physics, University of Western Australia, 35 Stirling Highway, Crawley WA 6009, Australia. Zeyu Kuang is with Nanjing University, 22 Hankou Road, Gulou District, Nanjing, Jiangsu, China, 210093. Philipp Haslinger and Holger Muller are with Department of Physics, University of California, Berkeley, California, US}}

\title{Next Generation of Phonon Tests of Lorentz Invariance using Quartz BAW Resonators}
\makeatletter
\def\ps@IEEEtitlepagestyle{
  \def\@oddfoot{\mycopyrightnotice}
  \def\@evenfoot{}
}
\def\mycopyrightnotice{
  {\footnotesize
  \begin{minipage}{\textwidth}
  \centering
  Copyright~\copyright~2017 IEEE. 
  \end{minipage}
  }
}

\begin{document}
\maketitle

\begin{abstract}

We demonstrate technological improvements in phonon sector tests of Lorentz Invariance that implement quartz Bulk Acoustic Wave oscillators. In this experiment, room temperature oscillators with state-of-the-art phase noise are continuously compared on a platform that rotates at a rate of order a cycle per second. The discussion is focused on improvements in noise measurement techniques, data acquisition and data processing. Preliminary results of the second generation of such tests are given, and indicate that SME coefficients in the matter sector can be measured at a precision of order $10^{-16}$ GeV after taking a years worth of data. This is equivalent to an improvement of two orders of magnitude over the prior acoustic phonon sector experiment. 
 
\end{abstract}

%\tableofcontents

\section{Introduction}

\IEEEPARstart{N}{owadays}, the Standard Model (SM) of particle physics is a widely accepted fundamental theory that classifies known elementary particle and forces. Despite its tremendous success in predicting and explaining relations between them, it leaves a significant amount of unanswered questions partly due to its incompatibility with General Relativity (GR). It is widely believed that the discovery of new physics beyond the SM and GR will help solve this problem. Local Lorentz invariance is a cornerstone of both SM and GR and one established way to look beyond these theories is to search of local Lorentz Invariance Violations (LIV). Test models that include such violations can be used to predict signals from a wide range of precision experiments, by far the most comprehensive is known as the Standard Model Extension (SME) proposed by Kosteleky and co-workers\cite{ColladayKostelecky,KosteleckyLane}. This extension includes SM and GR along with all possible terms that can violate Lorentz symmetry, i.e. by introducing anisotropies into different sectors of the SM and GR. The most well known of such anisotropies is the anisotropy of the speed of light widely investigated experimentally for more than a century\cite{MM2,MM3,KosteleckyMewesPRD,Stanwix,Hohensee,Schiller,Achim,Haeffner,Nagel}. Besides the theoretical description of the LIV terms, SME constitutes a framework allowing experimentalists to put bounds on various coefficients, which describe the 
Lorentz violating terms\cite{Coleman1999,Gomes,Bear,Altschul,Haeffner,KosteleckyTasson,Brown,WolfPRL2006}. The coefficients are grouped into four fundamental sectors dealing with light (photons), matter (electrons, protons, neutrons etc), neutrinos and gravity\cite{datatables}. Usually experiments or the analysis of pre-existing data (i.e. astrophysical or data from colliders) are implemented to be sensitive to a particular property in a particular sector. In this work, we perform precision measurements of oscillating masses of particles (or phonons), which constitute normal matter, i.e. electrons, protons and neutrons\cite{Lo:2016aa}.

The experiment described in this work is based on precision measurements of ultra-stable Bulk Acoustic Wave (BAW) quartz  oscillators. Although, frequency stability of these oscillators is surpassed by other frequency standards, i.e. atomic clocks, it is often that case that the sensitivity is limited by systematic effects and ability to maintain the experiment for very long times rather by intrinsic stability of the used sources. Thus, quartz oscillators, whose systematics have been studied for a few decades, make a very well understood, reliable and robust platform for this kind of measurements.  

%The proposed approach proved to be reliable and sensitive enough to push the current limits on neutron coefficients by a few orders of magnitude comparing to modern laboratory tests and astrophysics observation.

\section{Physical Principles and Previous Experiments}

The resonant frequency of mechanical resonators and BAW devices depend directly on the mode effective mass, and are thus widely utilised as precision mass sensors in many engineering, chemical and medical applications\cite{QCM1}. In principle, the resonant frequency depends not only on external loadings but also on variation of its intrinsic mass and thus on the inertial masses of composing particles. So, by modulating these masses, one modulates phonon mode frequencies that can be measured using precision frequency measurement techniques. For putative LIV in the matter sector of the SME, modulations of the internal masses of elementary particles are predicted to depend on the direction and boost velocity in space. Thus, the idea of the LIV test for ordinary matter particles in the 'phonon sector' reduces to measurements of frequency stability of mechanical resonators as a function of direction and boost in space, relative to some fixed reference frame. 

The phonon sector Lorentz invariance test setup is built around two frequency sources based on mechanical resonators. Ideally, the displacement vectors for both resonators should be orthogonal comparing internal masses of particles in two directions. As the setup rotates in space, e.g. with the rotating Earth, the difference between the two frequencies is modulated as proposed by the SME. Thus, by measuring the frequency stability of the pair the experiment is sensitive to the hypothetical SME LIV coefficients. Implementing this approach, the overall sensitivity is limited by the oscillator or resonator frequency fluctuations at time periods of order twice the rotation period, as well as all kinds of systematics associated with the rotation. Rotating the experiment effectively chops the signal, so long-term performance of the oscillators does not influence the measurements even when taking measurements for longer than one year. Since mechanical oscillators demonstrate the best frequency stability at relatively low integration times (less than a day), to achieve the best sensitivity, the oscillators have to be rotated on a turntable with a frequency corresponding to the integration time of order of the best stability. For this experiment this period is on the order of a second. This method has been used in the first generation of the phonon sector LIV tests\cite{Lo:2016aa}.
Among all frequency sources based on mechanical motion, quartz BAW oscillators provide the best frequency stability reaching below $10^{-13}$ between 1 and 10 seconds of integration time\cite{Salzenstein:2010aa} and low sensitivity to external and internal instability effects such as temperature, vibration, acceleration and ageing\cite{Vig:1991aa}. Temperature sensitivity of these devices is greatly suppressed with a double active oven control, and the impact of vibration and acceleration\cite{Filler:1988oa} is reduced by employing a resonator of specific cut according to a symmetrical arrangement\cite{Besson:1979aa}. Together with these facts, the overall simplicity and ability to sustain obtain optimum performance for very long periods of time make quartz Voltage Controlled Ovenized Crystal Oscillators (OCXO) an outstanding platform for phonon sector LIV tests. 

\section{Sensitivity Improvement}

Sensitivity of the first generation LIV test in the phonon sector\cite{Lo:2016aa} was limited by the frequency stability of the employed quartz oscillators ($\sim 10^{-12}$ of fractional frequency stability), relatively short observation times (120 hours) and the extraction of only one SME coefficient based on the noise power spectral density. In the new generation, these major issues have been addressed. Overall, the second generation system is improved by the following means:
\begin{itemize}
\item The frequency stability of mechanical oscillators is increased by introducing a pair of state-of-the-art quartz oscillators from Oscilloquartz with Allan deviation of $10^{-13}$ at 1 second:
\item Measurement time is increased by improving overall reliability of the setup and data acquisition and analysis techniques, allowing the collection and analysis of measurements of over one year time spans:
\item Data acquisition and noise measurement techniques are improved by using low noise measurement techniques and measurement devices;
\item Systematic effects related to mechanical tilts and jitter are reduced using an air-bearing turntable (RT300L Air Bearing from PI) with smaller tilts and rotation instabilities and a better quality rotating connector:
\item Long term stability is improved by putting the experiment into more stable and quiet environment with key parameters being monitored and controlled:
\item Data processing and fitting techniques are developed to better deal with large amounts of data and search for multiple SME coefficients at a variety of modulation frequencies.
\end{itemize}
 
\subsection{System Architecture}

\begin{figure}[htbp]
\centering
\includegraphics[width=1\columnwidth]{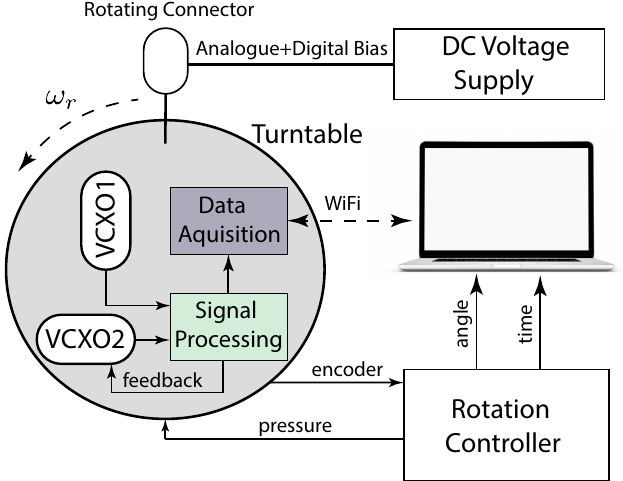}
\caption{Schematic of the rotating experimental setup. }
\label{setup}
\end{figure}

The main challenge of any rotating experiment is related to the need to supply and collect bias voltages, signals and data onto and from the rotating setup. In the current experiment these connections are organised as follows: both oscillators are placed on the rotating table, generated signals are processed via Phased Locked Loop (PLL) and an interferometer\cite{Ivanov:1998aa} on the turntable, only DC supply voltages to bias oscillators and amplifiers are supplied through the rotating connector, generated error signals are digitized on the table and transmitted to a stationary computer via a Wi-Fi module, the stationary computer controls the rotation, collects data from the digitizer and the rotation encoder. The overall system architecture is illustrated in Fig.~\ref{setup}.

The system is designed to maximally separate analogue and digital parts in order to reduce noise. The bottleneck of the system is the rotating connector which has a limited number of lines (8 closely situated liquid Mercury connections in a single body) and bandwidth ($<100$MHz). To avoid any signal corruption or crosstalk through the connector, only DC voltages are supplied through it. All data processing and digitisation is performed on top of the turntable. 

The data acquisition is controlled via a single Labview program that collects universal time, data from the error signals of the oscillator frequency comparison, time and angle from the rotation controller, environment temperature in a single data acquisition loop. This single loop approach guarantees the best matching between data collected on and off the turntable.

\subsection{Phase Noise Measurements from Two Oscillator Frequency Comparison}

\begin{figure}[htbp]
\centering
\includegraphics[width=1\columnwidth]{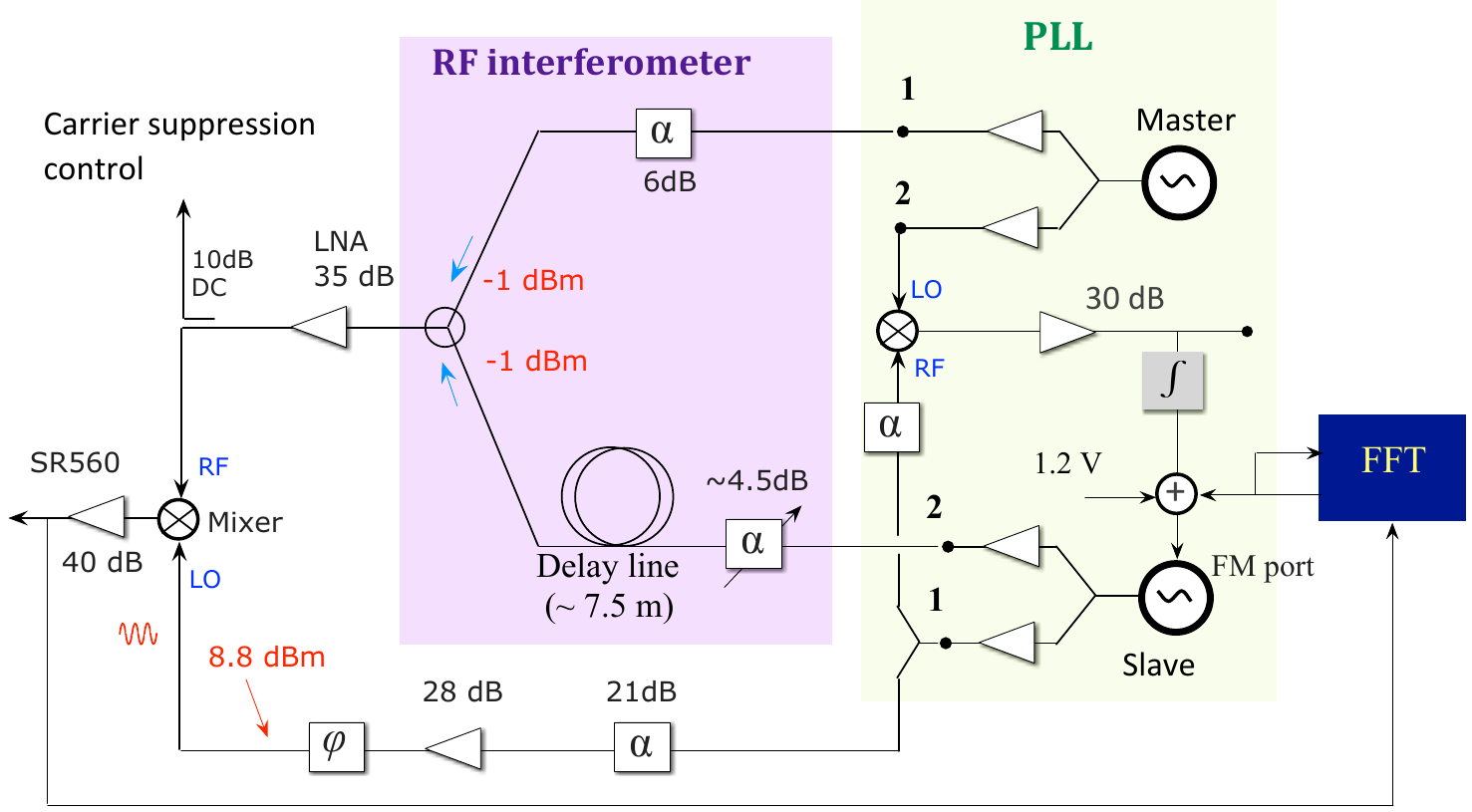}
\caption{Phase noise measurement setup implemented on the turntable. }
\label{measurement}
\end{figure}

\begin{figure}[htbp]
\centering
\includegraphics[width=0.9\columnwidth]{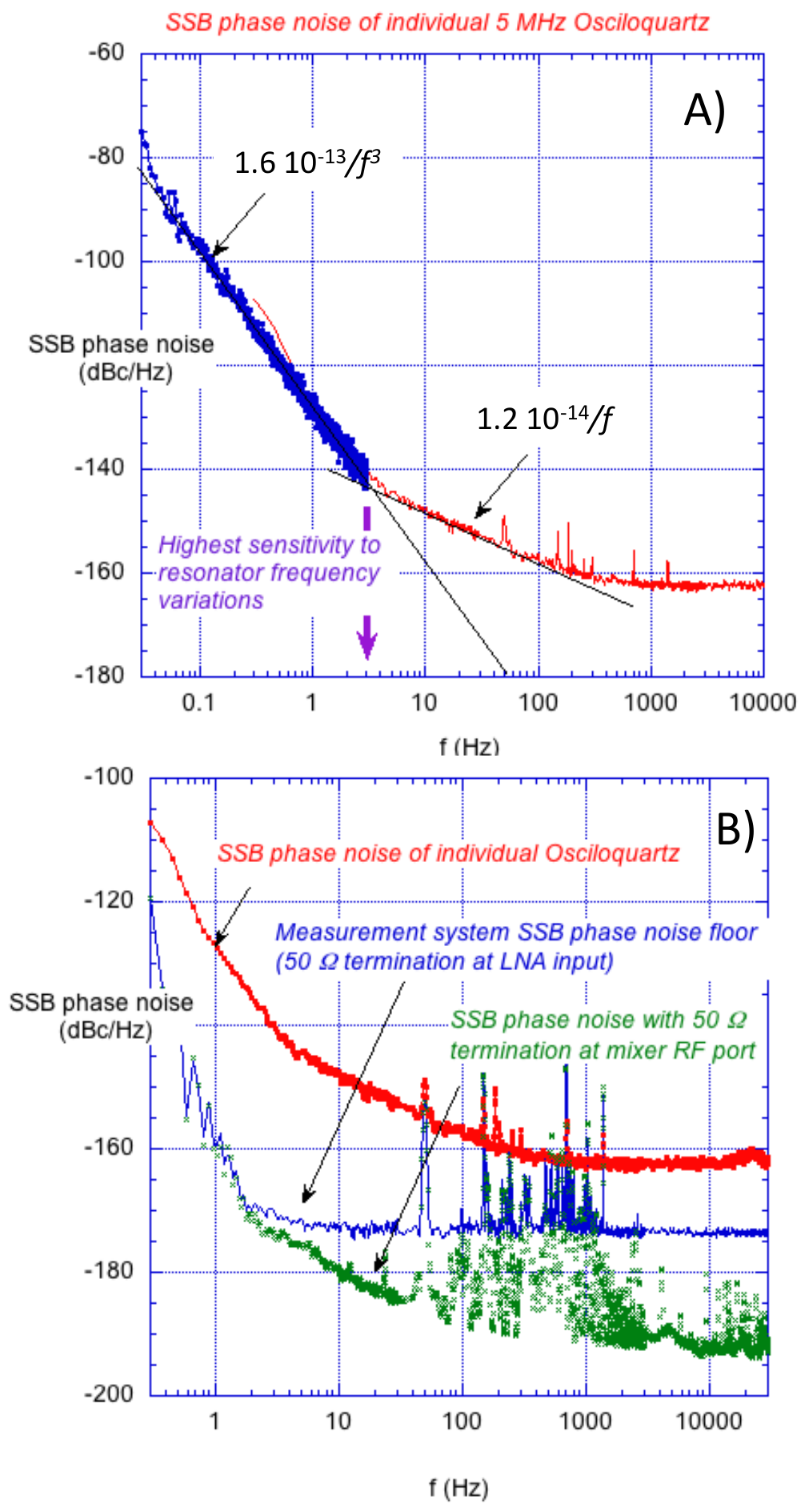}
\caption{Phase noise measurements of the system oscillator performance in the lab frame using (A) PLL and (B) interferometer. Results are consistent with the manufacturers stability measurements of a flicker floor of $10^{-13}$. Phase noise measurements at the lowest Fourier frequencies become inaccurate due to the finite duration of the data.}
\label{noiseres}
\end{figure}

As it has been described above, the task of LIV detection is reduced to frequency comparison measurements of two orthogonally orientated oscillators place on the turnable over large spans of time. In this work, two ultra-stable 5MHz OCXO are used. To measure their phase noise two approaches are implemented: one is based on a standard PLL technique, the other employs interferometric measurements. The overall schematic of the phase noise measurement and locking system is shown in Fig.~\ref{measurement}, while the results of the phase noise measurements of the two oscillators while stationary are shown in Fig.~\ref{noiseres}.

\begin{figure}[h]
\centering
\begin{subfigure}{.45\columnwidth}
	\centering
	\includegraphics[width=1\columnwidth]{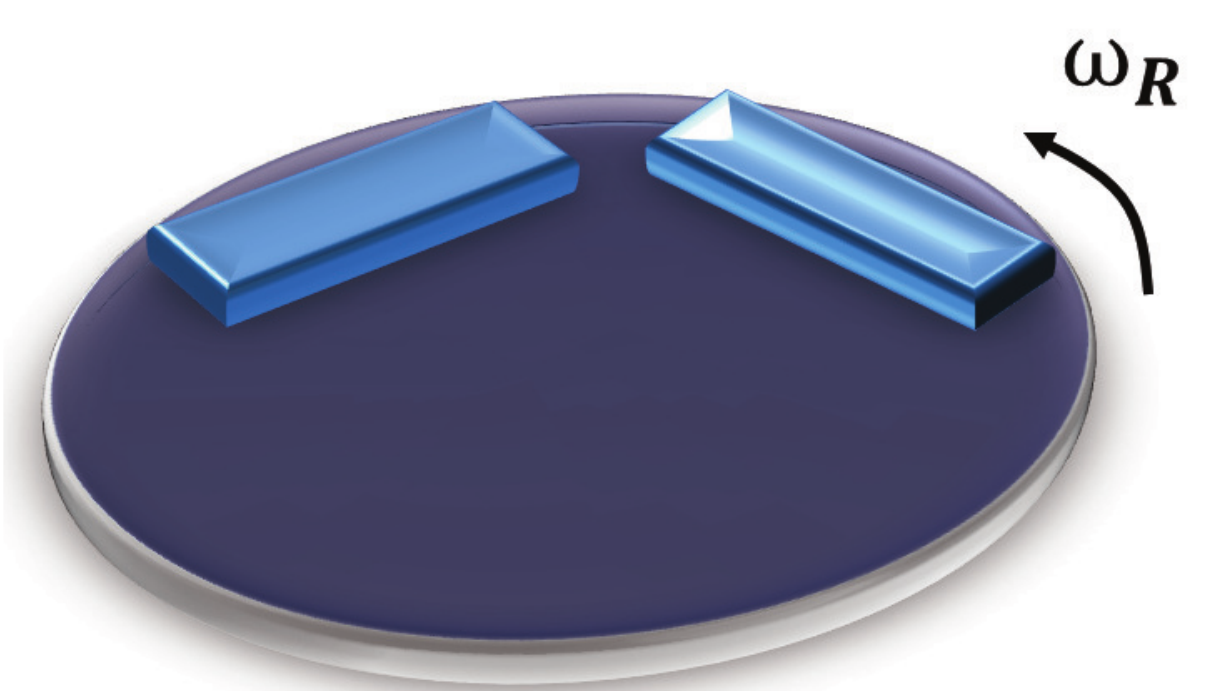}
	\caption{}
	%\caption{Two sapphire ocillators placed on a rotational table, rotating at frequency $\omega_R$}
\end{subfigure}%
\begin{subfigure}{.45\columnwidth}
	\centering
	\includegraphics[width=1\columnwidth]{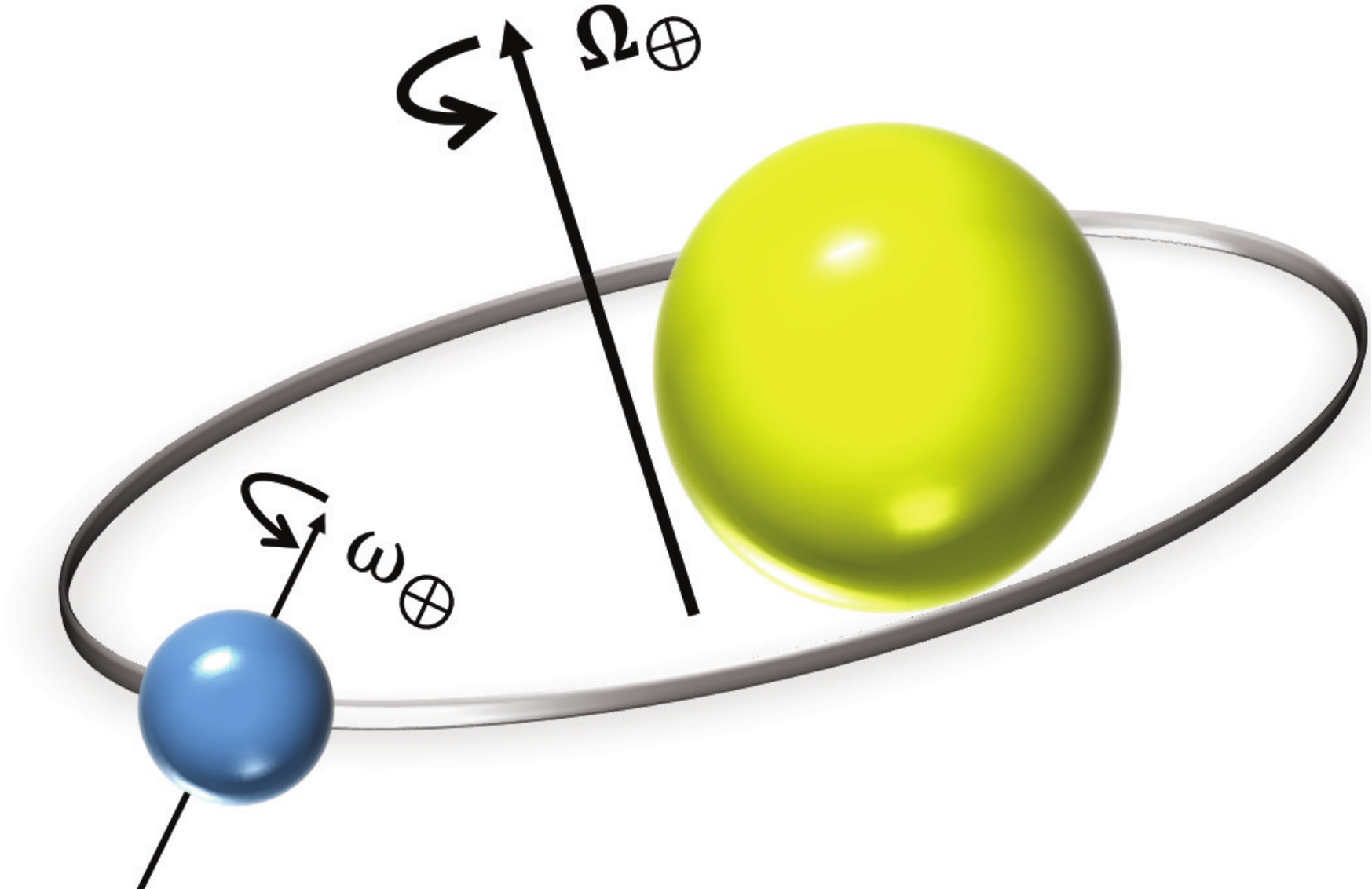}
	\caption{}
\end{subfigure}
\caption{Illustration of three frequency components: (a) Two orthogonal oscillators placed on a rotational table, rotating at frequency $\omega_R$. (b) Sidereal frequency $\omega_\oplus$ and annual frequency $\Omega_\oplus$.}
\label{ThreeOmega}
\end{figure}

\begin{figure}[h]
	\centering
	\includegraphics[width=1\columnwidth]{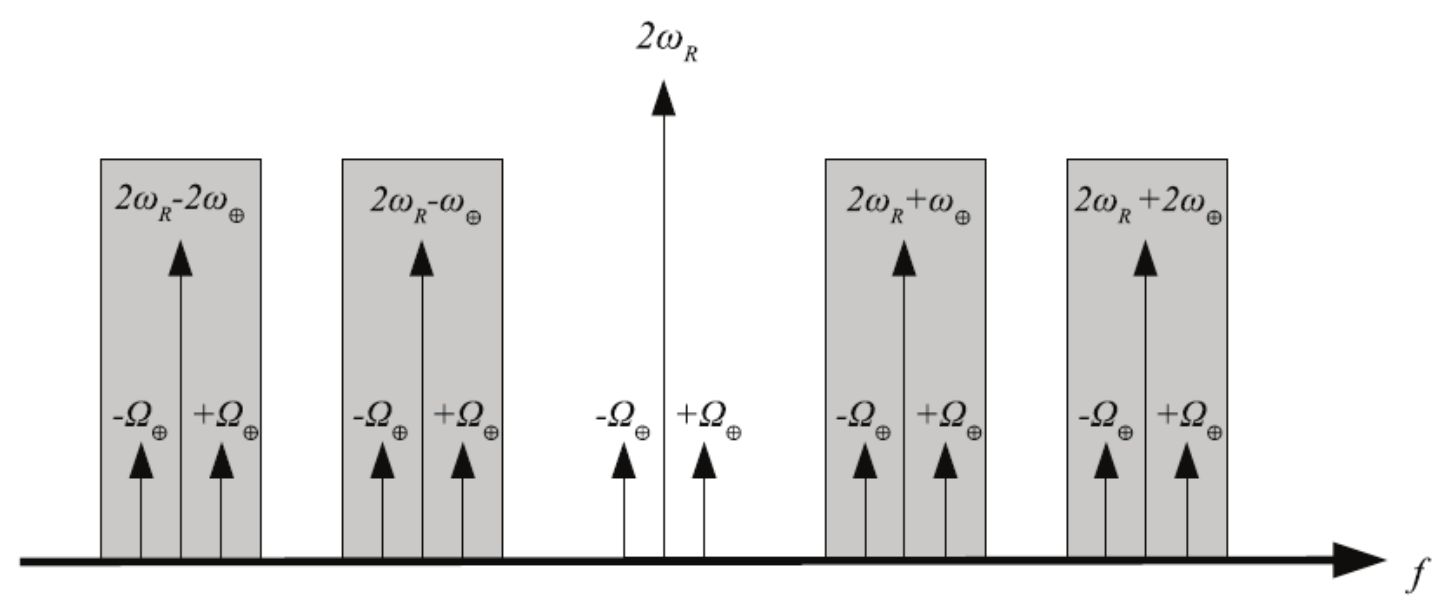}
	\caption{Illustration of frequency components of $\frac{\Delta f}{f}$caused by putative LIV coefficients}
	\label{Frequency stick diagram}
\end{figure}

\begin{table*}[h]
\begin{center}
\begin{tabular}{ |c|c|c| } 
	\hline
	$\omega_i$ (offset from $2\omega_R$)& $C_{\omega_i}$ & $S_{\omega_i}$ \\
	\hline\hline
	DC ($A$) & $-4\tilde{c}^T_Q\cos(2\theta)$ &  \\ 
	0 & $-4\sin^2\chi \tilde{c}^T_Q$ & 0 \\ 
	+ $\Omega_\oplus$ & $2\sin^2\chi (\cos\eta \tilde{c}^T_{TY} - 2 \tilde{c}^T_{TZ}\sin\eta )\beta_\oplus$ & $-2\sin^2\chi \tilde{c}^T_{TX}\beta_\oplus$ \\ 
	 - $\Omega_\oplus$ & $2\sin^2\chi (\cos\eta \tilde{c}^T_{TY} - 2 \tilde{c}^T_{TZ}\sin\eta )\beta_\oplus$ & $2\sin^2\chi \tilde{c}^T_{TX}\beta_\oplus$  \\ 
	+ $\omega_\oplus$ - $\Omega_\oplus$ & $2(1+\cos \chi)\tilde{c}^T_{TX}\sin\eta\sin\chi\beta_\oplus$ & \makecell{$2(1+\cos \chi)\sin\chi\beta_\oplus\times$ \\ $(\tilde{c}^T_{TZ}+\cos\eta \tilde{c}^T_{TZ} + \tilde{c}^T_{TY}\sin\eta)$} \\ 
	- $\omega_\oplus$ + $\Omega_\oplus$ & $2(\cos \chi-1)\tilde{c}^T_{TX}\sin\eta\sin\chi\beta_\oplus$ & \makecell{$2(\cos \chi-1)\sin\chi\beta_\oplus\times$ \\ $(\tilde{c}^T_{TZ}+\cos\eta \tilde{c}^T_{TZ} + \tilde{c}^T_{TY}\sin\eta)$} \\ 
	 + $\omega_\oplus$ & $-4(1+\cos \chi)\tilde{c}^T_{Y}\sin\chi $ & $-4(1+\cos \chi)\tilde{c}^T_{X}\sin\chi$ \\
	 - $\omega_\oplus$ & $4(\cos \chi-1)\tilde{c}^T_{Y}\sin\chi $ & $4(\cos \chi - 1)\tilde{c}^T_{X}\sin\chi$ \\
	+ $\omega_\oplus$ + $\Omega_\oplus$ & $2(1+\cos\chi)\tilde{c}^T_{TX}\sin\eta\sin\chi\beta_\oplus$ & \makecell{$2(1+\cos \chi)\sin\chi\beta_\oplus\times$\\$[(-1+\cos\eta)\tilde{c}^T_{TZ}+\sin\eta \tilde{c}^T_{TY}]$}  \\
	 - $\omega_\oplus$ - $\Omega_\oplus$ & $2(\cos\chi-1)\tilde{c}^T_{TX}\sin\eta\sin\chi\beta_\oplus$ & \makecell{$2(1-\cos \chi)\sin\chi\beta_\oplus\times$\\$[(-1+\cos\eta)\tilde{c}^T_{TZ}+\sin\eta \tilde{c}^T_{TY}]$}  \\
	+ 2$\omega_\oplus$ - $\Omega_\oplus$ & $(1+\cos\eta)(1+\cos\chi)^2\tilde{c}^T_{TY}\beta_\oplus$ & $-(1+\cos\eta)(1+\cos\chi)^2\tilde{c}^T_{TX}\beta_\oplus$ \\ 
	 - 2$\omega_\oplus$ + $\Omega_\oplus$ & $(1+\cos\eta)(-1+\cos\chi)^2\tilde{c}^T_{TY}\beta_\oplus$ & $(1+\cos\eta)(-1+\cos\chi)^2\tilde{c}^T_{TX}\beta_\oplus$ \\ 
	 + 2$\omega_\oplus$ & $2\tilde{c}^T_-(1+\cos\chi)^2$ & $2(1+\cos\chi)^2\tilde{c}^T_Z$ \\
	- 2$\omega_\oplus$ & $2(\cos\chi-1)^2\tilde{c}^T_-$ & $-2(\cos\chi-1)^2\tilde{c}^T_Z$ \\
	 + 2$\omega_\oplus$ + $\Omega_\oplus$ & $(\cos\eta-1)(1+\cos\chi)^2\tilde{c}^T_{TY}\beta_\oplus$ & $(1-\cos\eta)(1+\cos\chi)^2\tilde{c}^T_{TX}\beta_\oplus$ \\
	 - 2$\omega_\oplus$ - $\Omega_\oplus$ & $(\cos\eta-1)(-1+\cos\chi)^2\tilde{c}^T_{TY}\beta_\oplus$ & $(\cos\eta-1)(-1+\cos\chi)^2\tilde{c}^T_{TX}\beta_\oplus$ \\
	\hline
\end{tabular}
\end{center}
\caption{Relation between SME neutron $c$ coefficients and frequency components of Eqn. (\ref{FrequencyCoupling}), which are also pictorially shown in Fig.\ref{Frequency stick diagram}. Note, these coefficients were originally derived in \cite{Lo:2016aa}, they are presented here again with a few small typographical errors fixed. Here $\chi$ is the colatitude of the lab, $\eta$ is the declination of the Earth's orbit relative to the sun centered frame and $\beta_\oplus$ is the boost of the lab with respect to the sun centered frame.}
\label{tb:SMErelation}
\end{table*}

For the rotating setup the voltage signals from the PLL and the interferometer are digitized on the turntable using a National Instruments high speed digitizer. The signals are sampled at $1.6$ kHz and averaged over 50 samples to achieve a balance between the amount of data and noise levels. System operation over month time scales demonstrate high reliability of the oscillator as well as the PLL and the interferometer. While the PLL stays always firmly locked, the interferometer carrier suppression may vary. Despite this, the interferometer stays operational over months of non-stop measurements and remains phase sensitive with a constant voltage to phase conversion and with sufficient sensitivity to measure the oscillator phase noise. Such results are achieved due to high environmental stability. 

Although, the interferometer is capable to higher sensitivity of the phase noise measurements, for these particular LIV tests only a small frequency range around twice the rotational frequency is important. Typically, in this frequency range ($1-5$Hz) both measurement techniques give the same results as both are sensitive enough to measure the oscillator phase noise. Thus, the implementation of the interferometer in the future runs is not necessary for this particular experiment. However the addition of the redundant measurement system has helped to disentangled the influence of the systematic signal through the quartz oscillators and the measurement system, as both systems show the same spurious signal-noise ratios. For example, the magnetic field to voltage conversion can be attributed to the OCXOs rather than to the measurement apparatus.

\section{Data Analysis}

Besides rotation in the laboratory, If we assume that Lorentz violation exists, then the frequencies of the two resonators would differ by $\Delta f$. The fractional frequency difference, $\frac{\Delta f}{f}$, has three major frequency components, $2\omega_R$, $\omega_\oplus$, and $\Omega_\oplus$. Here $\omega_R$ is the rotational frequency of the experimental table, $\omega_\oplus$ is the sidereal frequency and $\Omega_\oplus$ is the annual frequency, as shown in Figure \ref{ThreeOmega}.

It has already been shown in \cite{Lo:2016aa} that the LIV coefficients in the Standard Model Extension (SME) test theory, which are constant in the sun-centred frame, will cause coherent frequency modulations with respect to the laboratory frame. This is calculated by undertaking the Lorentz transformations of rotations and boosts experienced by the experiment with respect to the sun-centred frame due to rotation in the lab and sidereal and annual orbit rotations. Thus, the expected frequency shift is given by;
\begin{equation}
\frac{\Delta f}{f} = \frac{1}{8}(A +  \sum_{i} [S_i \sin((2\omega_R+\omega_i)T_\oplus) + C_i \cos((2\omega_R+\omega_i)T_\oplus)])
\label{FrequencyCoupling}
\end{equation}
Here, the $i^{th}$ possible putative frequency shifts occurs at $2\omega_R+\omega_i$,  $T_\oplus$ is the local sidereal time defined as the time from the vernal equinox in the year 2000 and $A$ is the $DC$ shift. The frequency components are illustrated diagrammatically in Figure \ref{Frequency stick diagram}. We list a similar table to that as published in \cite{Lo:2016aa} in Tab.\ref{tb:SMErelation}, however we point out there are some slight differences due to typographical errors in the one presented in \cite{Lo:2016aa}.

\subsection{Demodulated Least Square Method}

\begin{table*}
\begin{center}
\begin{tabular}{|c|c|c|}
\hline
$\omega_i$ & $C_{C,\omega_i}$  & $C_{S,\omega_i}$ \\
\hline
$0$ & $-4\sin^2\chi \tilde{c}^T_Q$ & 0 \\
$\Omega_\oplus$ & $4\sin^2\chi(\cos\eta \tilde{c}^T_{TY}-2\tilde{c}^T_{TZ}\sin\eta)\beta_\oplus$ & $-4\sin^2\chi \tilde{c}^T_{TX}\beta\oplus$ \\
$\omega_\oplus - \Omega_\oplus$ & $4\cos\chi \tilde{c}^T_{TX}\sin\eta\sin\chi\beta_\oplus$ & \makecell{$4\cos\chi(\tilde{c}^T_{TZ}+\cos\eta \tilde{c}^T_{TZ}+\tilde{c}^T_{TY}\sin\eta)\times$\\$\sin\chi\beta_\oplus$} \\
$\omega_\oplus$ & $-8\tilde{c}^T_Y\sin\chi$ & $-8\cos\chi \tilde{c}^T_X \sin\chi$ \\
$\omega_\oplus + \Omega_\oplus$ & $4\cos\chi \tilde{c}^T_{TX}\sin\eta\sin\chi\beta_\oplus$ & \makecell{$4[(-1+\cos\eta)\tilde{c}^T_{TZ}+\cos\chi \tilde{c}^T_{TY}\sin\eta]\times$\\$\sin\chi\beta_\oplus$}\\
$2\omega_\oplus - \Omega_\oplus$ & $2(1+\cos\eta)(1+\cos^2\chi)\tilde{c}^T_{TY}\beta_\oplus$ & $-2(1+\cos^2\chi)(1+\cos\eta)\tilde{c}^T_{TX}\beta_\oplus$ \\
$2\omega_\oplus$ & $4\tilde{c}^T_-(\cos^2\chi+1)$ & $4(1+\cos^2\chi)\tilde{c}^T_Z$ \\
$2\omega_\oplus + \Omega_\oplus$ & $2(1+\cos^2\chi)(\cos\eta-1)\tilde{c}^T_{TY}\beta_\oplus$ & $2(1-\cos\eta)(1+\cos^2\chi)\tilde{c}^T_{TX}\beta_\oplus$ \\
\hline
\hline
$\omega_i$ & $S_{C,\omega_i}$  & $S_{S,\omega_i}$ \\
\hline
$0$ & $0$ & 0 \\
$\Omega_\oplus$ & $0$ & $0$ \\
$\omega_\oplus - \Omega_\oplus$ & $4(\tilde{c}^T_{TZ}+\cos\eta \tilde{c}^T_{TZ} + \tilde{c}^T_{TY}\sin\eta)\sin\chi\beta_\oplus$ & $-4\tilde{c}^T_{TX}\sin\eta\sin\chi\beta_\oplus$ \\
$\omega_\oplus$ & $-8\tilde{c}^T_X\sin\chi$ & $8\cos\chi \tilde{c}^T_Y \sin\chi$ \\
$\omega_\oplus + \Omega_\oplus$ & $4[\cos\chi(-1+\cos\eta)\tilde{c}^T_{TZ}+\tilde{c}^T_{TY}\sin\eta]\sin\chi\beta_\oplus$ & $-4\tilde{c}^T_{TX}\sin\eta\sin\chi\beta_\oplus$\\
$2\omega_\oplus - \Omega_\oplus$ & $-4(1+\cos\eta)\cos\chi \tilde{c}^T_{TX}\beta_\oplus$ & $-4(1+\cos\eta)\cos\chi \tilde{c}^T_{TY}\beta_\oplus$ \\
$2\omega_\oplus$ & $8\tilde{c}^T_Z\cos\chi$ & $-8\tilde{c}^T_-\cos\chi$ \\
$2\omega_\oplus + \Omega_\oplus$ & $4(1-\cos\eta)\cos\chi \tilde{c}^T_{TX}\beta_\oplus$ & $-4\cos\chi(\cos\eta-1)\tilde{c}^T_{TY}\beta_\oplus$ \\\hline
\end{tabular}
\end{center}
\caption{Relationship between DLS parameters and SME neutron $c$ coefficients from Eqns. (\ref{Eq15}) and (\ref{Eq16}), which is shown pictorially in Fig. \ref{Fig6}. Here $\chi$ is the colatitude of the lab, $\eta$ is the declination of the Earth's orbit relative to the sun centered frame and $\beta_\oplus$ is the boost of the lab with respect to the sun centered frame.}
\label{DLS_SME}
\end{table*}

Many experiments that search for putative LIV coefficients apply the technique of least square fitting \cite{Wolf2003,Wolf2004,Stanwix,Tobar2010,Tobar2013}. However, for experiments that compare oscillators, minimum frequency instabilities typically occur on time scales of the order of 1 to 100 seconds\cite{Abgrall2016}, so rotating the experiment can significantly enhance the precision of the experiment, compared to relying on the Earth rotation. When combined with rotation in the lab, the technique of Demodulated Least Squares (DLS) becomes a favourable technique\cite{Stanwix2006,Meuller2007,Tobar2009,Parker2011}. This is because the data files can become quite large, when taking such data over a period of one year. For example, if an experiment relies on sidereal rotation data maybe averaged over thousands of seconds to resolve a sidereal period resulting in the order of $10^{4}$ data points to search for the required frequency modulations. However, to attain maximum sensitivity we necessarily rotate our experiment on the order of 1 second, and therefore must take data with a measurement time of order 0.1 seconds, which leads to a requirement of searching for LIV with $10^{8}$ to $10^{9}$ data points. Thus, implementing the least squares method to extract the required coefficients from such a long set of data will become lengthy. We found that the DLS technique not only is quicker, but can extract parameters with better signal to noise ratio with respect to ordinary least squares (OLS), but in contrast requires a two stage process.

In the first stage, we effectively demodulate the $2\omega_R$ components using OLS to create a demodulated data set, while in the second stage, we extract the expected frequency components from the created data set. Compared to using OLS the DLS technique decreases the time necessary to process the data. This is because most of the data is averaged in the first stage over a finite number of rotations at the largest frequency component, $2\omega_R$. It has been shown there is an optimum number of rotations, which will balance of the narrow band systematic noise due to rotation and the broad band electronic noise in the system\cite{Stanwix2006}.

Equation (\ref{FrequencyCoupling}) can be rewritten as a function of $2\omega_R$ as;
\begin{equation}
\frac{\Delta f}{f} = \frac{1}{8}(A + S(T_\oplus)\sin(2\omega_RT_\oplus) + C(T_\oplus)\cos(2\omega_RT_\oplus)).
\label{Eq14}
\end{equation}
In this rearrangement, $S(T_\oplus)$ and $C(T_\oplus)$ contains rest of the frequency components;
\begin{align}
S(T_\oplus) &= S_0 + \sum_iS_{s,i}\sin(\omega_iT_\oplus) + S_{c,i}\cos(\omega_iT_\oplus) \label{Eq15}\\
C(T_\oplus) &= C_0 + \sum_iC_{s,i}\sin(\omega_iT_\oplus) + C_{c,i}\cos(\omega_iT_\oplus). \label{Eq16}
\end{align}
\begin{figure}[h]
\center
\includegraphics[width=0.75\columnwidth]{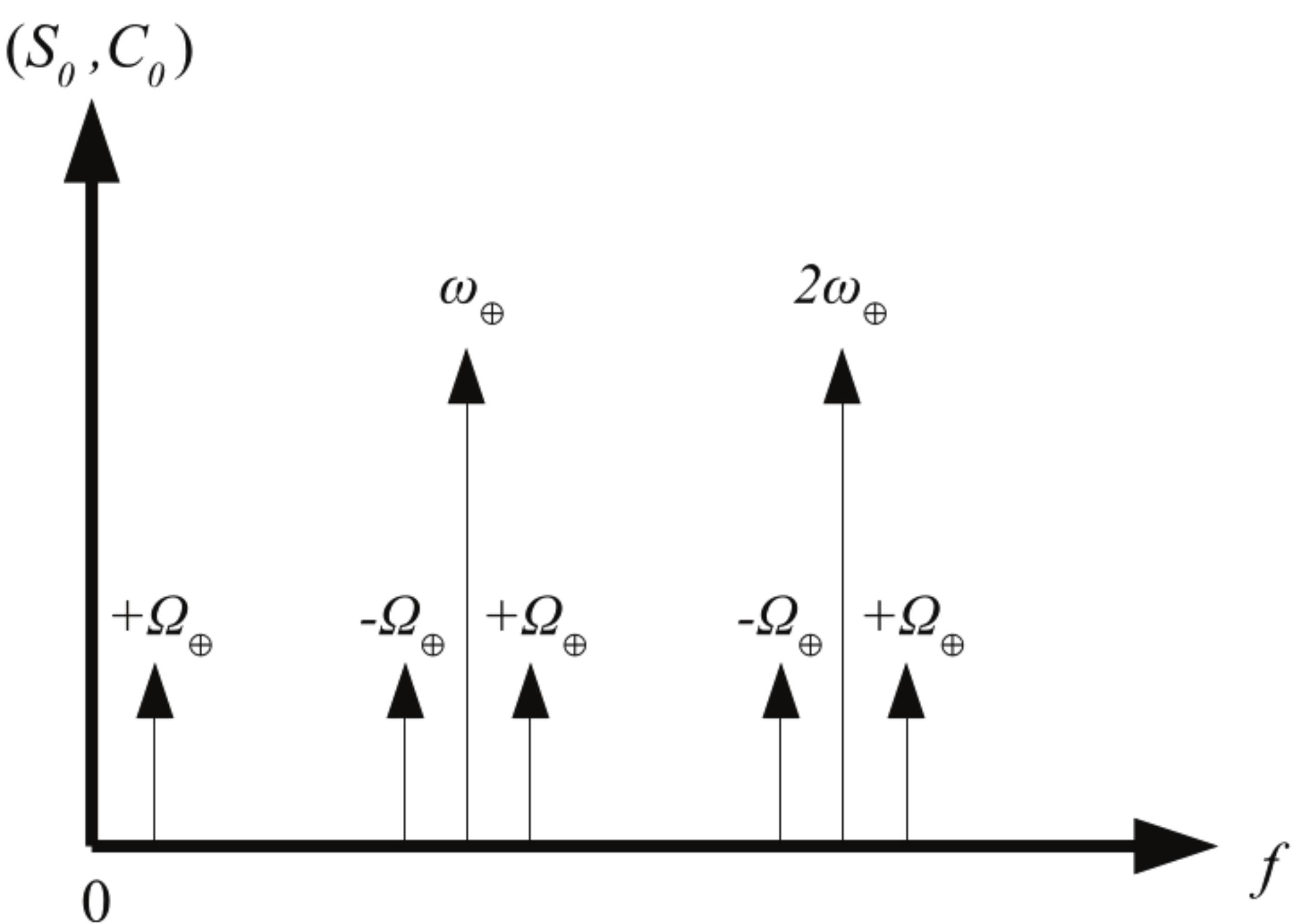}
\caption{Illustration of frequency components of $\frac{\Delta f}{f}$caused by putative LIV coefficients once the frequency is demodulated.}
\label{Fig6}
\end{figure}
Each demodulated frequency, $\omega_i$, corresponds to each frequency component in Figure \ref{Fig6}. Compared to Figure \ref{Frequency stick diagram}, Figure \ref{Fig6} only have positive $\omega_i$. This is because the $-\omega_i$ and $\omega_i$ components are added together when demodulating. The corresponding demodulated data files are now similar to an experiment that uses no active rotation in the laboratory and the least square method may be used to find corresponding putative LIV coefficients. Previously, the sensitivity to neutron coefficients in the SME were calculated for only the case in Figure \ref{ThreeOmega}\cite{Lo:2016aa} at only $2\omega_R$. Here we calculate the relation between the demodulated frequency amplitude and the neutron SME coefficients, which are shown in Table \ref{DLS_SME}. We can use these relations to calculate SME coefficients from the least square fitted parameters of the second stage of the DLS.

\subsection{Data extraction and pre-processing}

At each time $t$, we effectively measure the difference of the two resonance frequencies by measuring the voltage, $V(t)$, at the output of the PLL or Mixer. We record the time $t$ and the angle of the rotational table $\phi_R(t)$. From $V(t)$, $\phi_R(t)$, and $t$, we may search for SME coefficients through least square fitting of each frequency component. In our experiment, the data has been divided into four runs; run 8, run 9, run 10, and run 11, as shown in Table \ref{charOfeachRun}. They start at different times and in general have different rotation speed $\omega_R$, which is defined as follows in Equation \ref{eq5}:
\begin{equation}
\omega_R = \frac{d\phi_R}{dt}
\label{eq5}
\end{equation}
We analyze individually the seperate runs to study the effect of modifications undertaken between each experimental run.

\begin{table}[h]
\begin{center}
\begin{tabular}{|c|c|c|c|c|}
\hline
 & run 8 &  run 9 &  run 10 &  run 11 \\
\hline
Date & 16/03/17 & 29/03/17 & 23/03/17 & 16/03/17 \\
\hline
Time & 5:41:09 am  & 8:06:14 am & 1:30:23 pm  & 10:57:57 am \\
\hline
$\omega_R$ & 360$^{\circ}/s$ & 360$^{\circ}/s$ & 420$^{\circ}/s$ & 320$^{\circ}/s$ \\
\hline
Days & 13 & 54 & 23 & 63  \\
\hline
\end{tabular}
\end{center}
\caption{Characteristic of each data run. Here, the Date and Time are the starting time for each run, recorded in UTC format. Different runs had different $\omega_R$. Run 8 and 9 had the same $\omega_R$ and were recorded continuously. Days indicates the time span of each run.} 
\label{charOfeachRun}
\end{table}

\begin{figure}[h]

\center
\includegraphics[width=1\columnwidth]{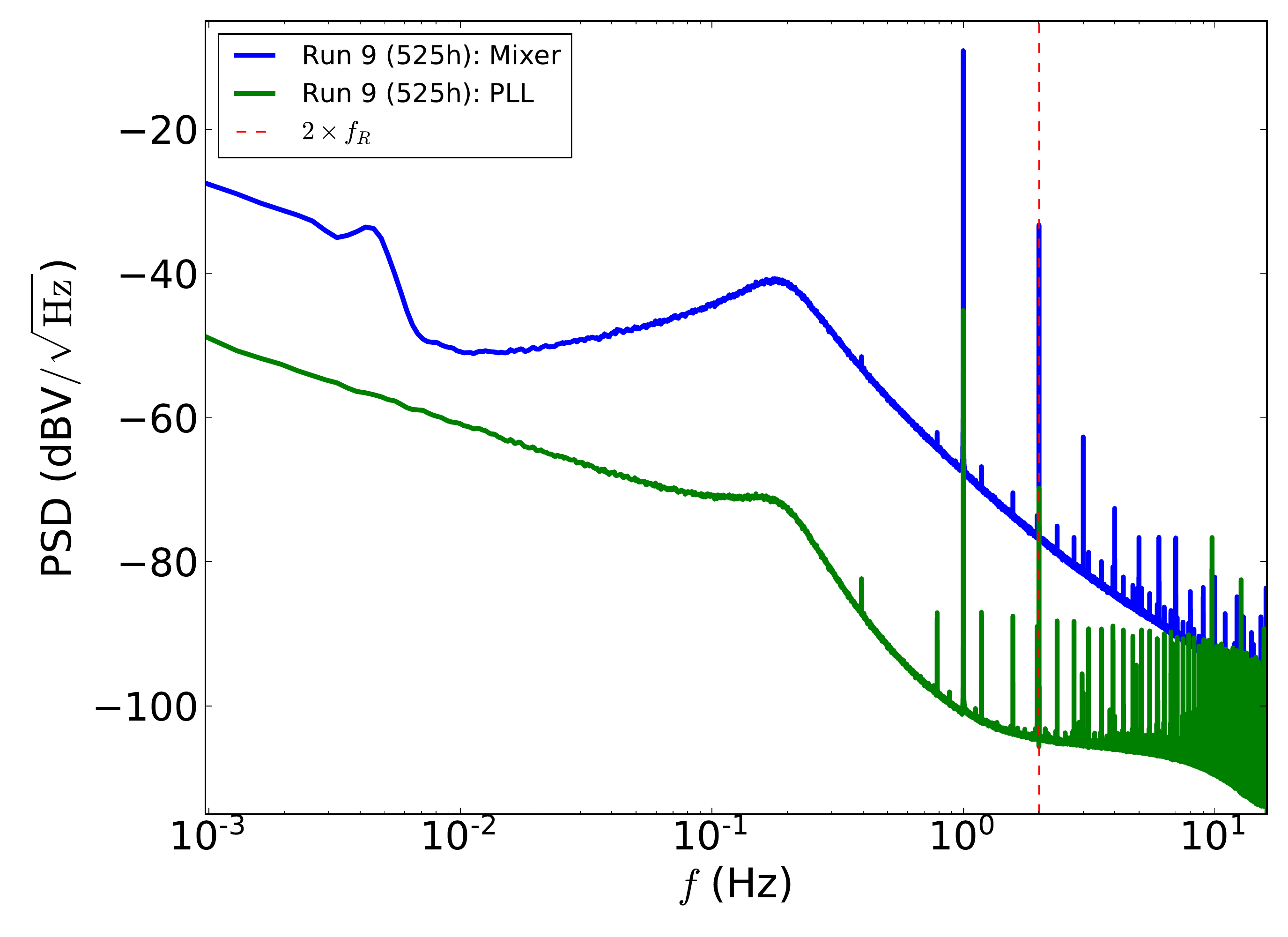}
\caption{Power Density Spectrum of the voltage measured from Mixer and PLL in run 9.}
\label{Fig3}
\end{figure}

\begin{figure}[h]
\center
\includegraphics[width=1\columnwidth]{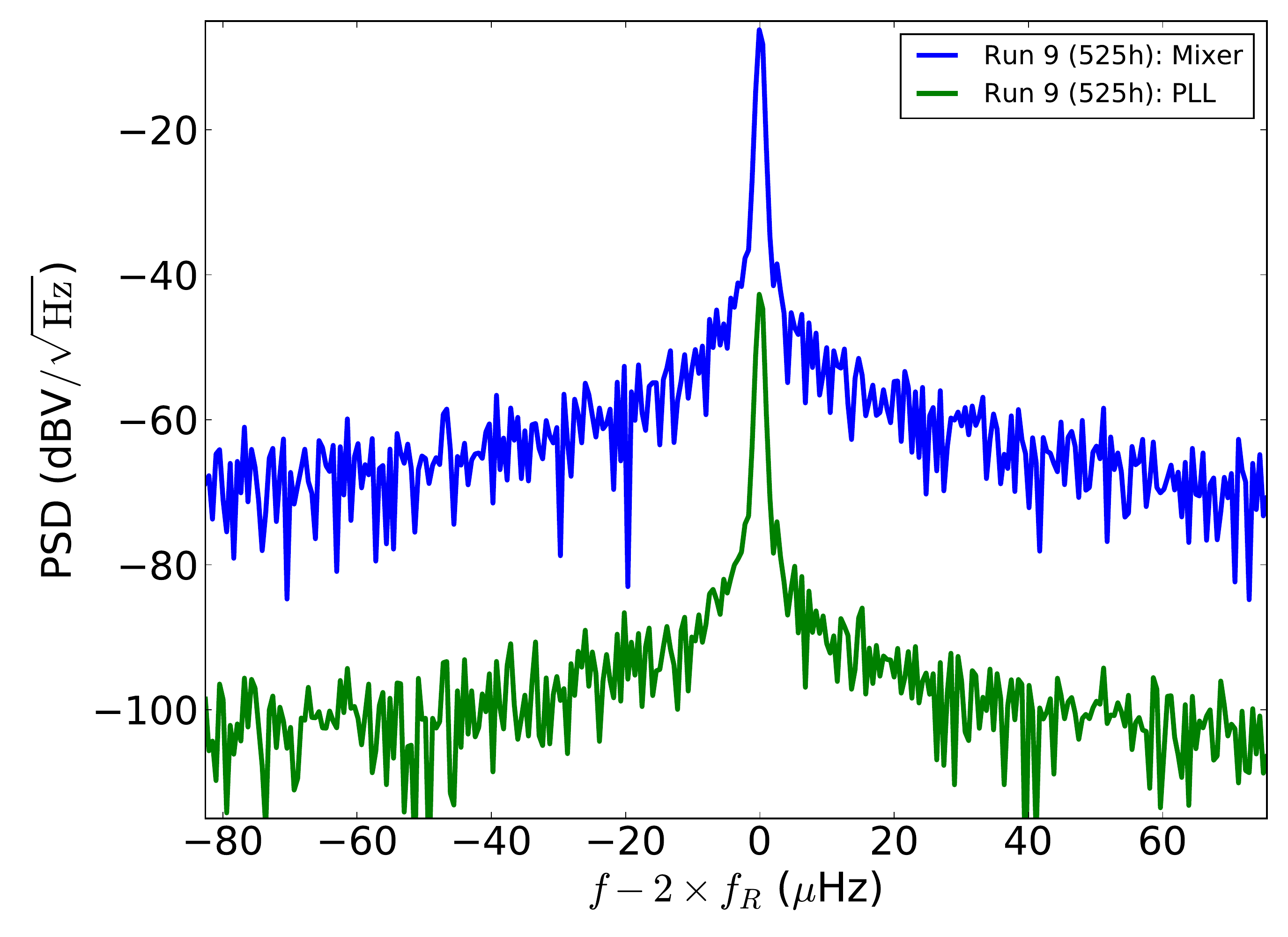}
\caption{Power Density Spectrum of the voltage in run 9, zoomed in at twice the rotation frequency}
\label{Fig4}
\end{figure}

The fractional frequency difference of the two oscillators was recorded from both the PLL and the output Mixer of the interferometer, and we calculated the Power Spectrum Density (PSD) from the time series of voltage measured from these ports. Here we use run 9 as an example, we calculated its PSD from both PLL and Mixer, as shown in Figure \ref{Fig3}. We zoom in at twice the rotational frequency, as shown in Figure \ref{Fig4}. Comparing the data recorded by PLL and Mixer, we can see that the frequency dependence of the PSD is similar. The difference in magnitude is mainly due to the different calibration factor. It was confirmed that the difference in sensitivity in both channels was negligible from start to finish of the data extraction and analysis procedure, so in the end we only present results from the voltage output of the PLL. There were some different spurious signals at other frequencies when comparing both systems, but this had no impact on the sensitivity near twice the rotation frequency.

There are two things we need to convert before searching for LIV. Firstly, we need to convert the voltage to fractional frequency, which will depend on the Fourier frequency offset since the two oscillator measurement is proportional to phase. Secondly, we need to convert local time tag to the time tag in the sun-centered frame \cite{datatables}. To convert the voltage from Mixer to the fractional frequency of two oscillators. We define the linear conversion factor $C_f$ so that:
\begin{equation}
\frac{\Delta f}{f} = C_f V
\end{equation}
where $V$ is the voltage from the Mixer. $C_f$ is linearly proportional to $\omega_R$ and since the runs differ by their rotational speed, any putative signals would occur close to $2\times f_R$, thus the measurements would have different $C_f$ due to appearing at a different Fourier frequency, as shown in Table 3. Furthermore, we need to convert the local time, $t$, to the local sidereal time in the sun centered frame, $T_\oplus$, by adding an offset $t_0$ \cite{datatables}.
\begin{equation}
T_\oplus = t + t_0
\end{equation}
The offset changes the local frame to the sun centered frame. Because we need to measure the physical quantity in an inertial frame, we set the sun-centered frame as the standard frame for our local coordinates. This means that all the physical quantity should be calculated from this standard frame. The sidereal phase (phase with respect to the sidereal rotation of the Earth), $\phi_\oplus$, is one of the quantities we are interested in:
\begin{equation}
\phi_\oplus = \omega_\oplus T_\oplus.
\end{equation}
If our local frame coincides with the standard frame when the experiment just started, then $t=0$ and $\phi_\oplus = 0$, and $t=T_\oplus$. However, this is highly unlikely, in general it is necessary to add an offset, $t_0$, to the local time tag $t$ to make sure the sidereal phase is measured in the standard frame. Since the different runs started at a different times, their $t_0$ offsets are different, as shown in Table \ref{Tab3}. Note the offsets are calculated after subtracting multiple values of $2\pi$ with respect to the vernal equinox in the year 2000, so the beginning of run 8 starts as close as possible to the value of $t_0 = 0$, without becoming negative.

\begin{table}[h]
\begin{center}
\begin{tabular}{|c|c|c|c|c|}
\hline
 & run 8 &  run 9 &  run 10 &  run 11 \\
\hline
$t_0(s)$&4055.88& 1135960.88 & 5907409.88 & 7971863.88 \\
\hline
$C_f(1/V)\times 10^{8}$ & 6.65726 & 6.65726 & 7.76803 & 5.91756\\
\hline
\end{tabular}
\end{center}
\caption{Pre-processing parameters, $t_0$ and $C_f$ for each run, $t_0$ translates the time tag to local sidereal time, while $C_f$ calibrates the Mixer Voltage to fractional frequency for the selected rotation frequency.}
\label{Tab3}
\end{table}

\section{Preliminary Results}

\begin{figure}[h]
	\center
	\includegraphics[width=1\columnwidth]{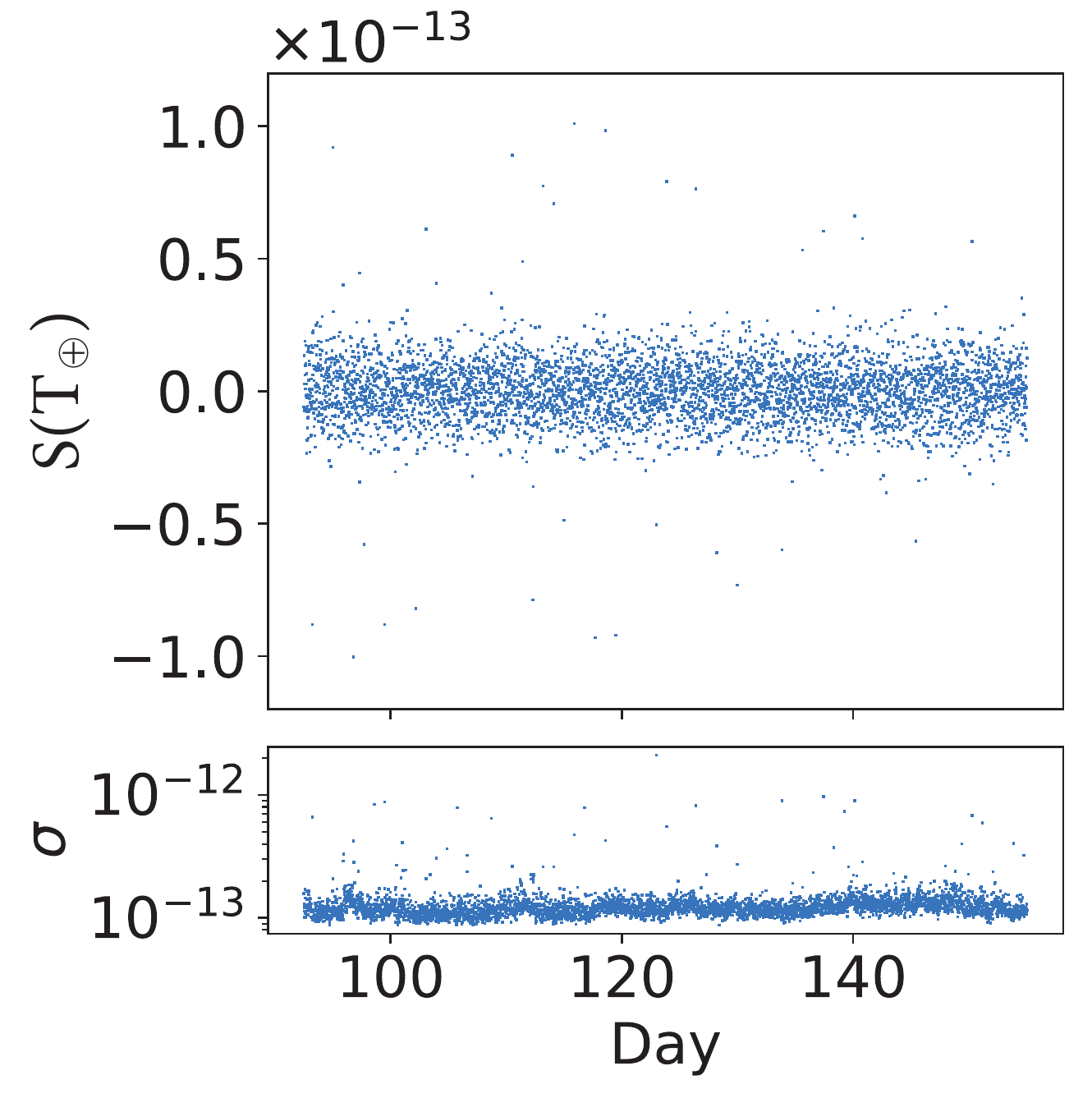}
	\caption{Data values for $S(T_\oplus)$ as a function of time in days for $N_r=1000$ for run 11, here $\sigma$ is the standard deviation of $S(T_\oplus)$.}
	\label{SandCvsDay}
\end{figure}

From the pre-processed data, we obtained data files with $\frac{\Delta_f}{f}$, $T_\oplus$, and $\phi_R$. During the first stage of the DLS, we seperate the processed data into several continuous subsets. The size of the subset was characterized by the number of rotations, $N_r$, chosen to optimise the signal to noise ratio. Implementing the OLS method for each subset, we extract values of $S(T_\oplus)$ and $C(T_\oplus)$ from Equation \ref{Eq14}. The time tag was then set as the averaged local sidereal time of the subset. For example, fitted values of $S(T_\oplus)$ from run 11 is shown in Figure \ref{SandCvsDay}.

The data files containing $S(T_\oplus)$ and $C(T_\oplus)$ were largely reduced in size compared to the original size. For example, we determined in run 11 that fitting over $N_r=1000$ rotations gave optimal signal to noise ratio, which in turn reduces the original data set by more than a factor of $10^4$. In the second stage of the DLS, we then fit the coefficients $S_{s,i}$,  $S_{c,i}$, $C_{s,i}$, and $C_{c,i}$ from Equation \ref{Eq15} and \ref{Eq16}, using the OLS method (it is also possible to use weighted least squares if the noise deviates far from white noise). For these preliminary results, we ignore the annual frequency components because the data set is too small to resolve them. In the future after a year of data has been taken it will be possible to put limits on all SME coefficients indicated in Table \ref{DLS_SME}.

\begin{figure}[h]
\center
\includegraphics[width=1\columnwidth]{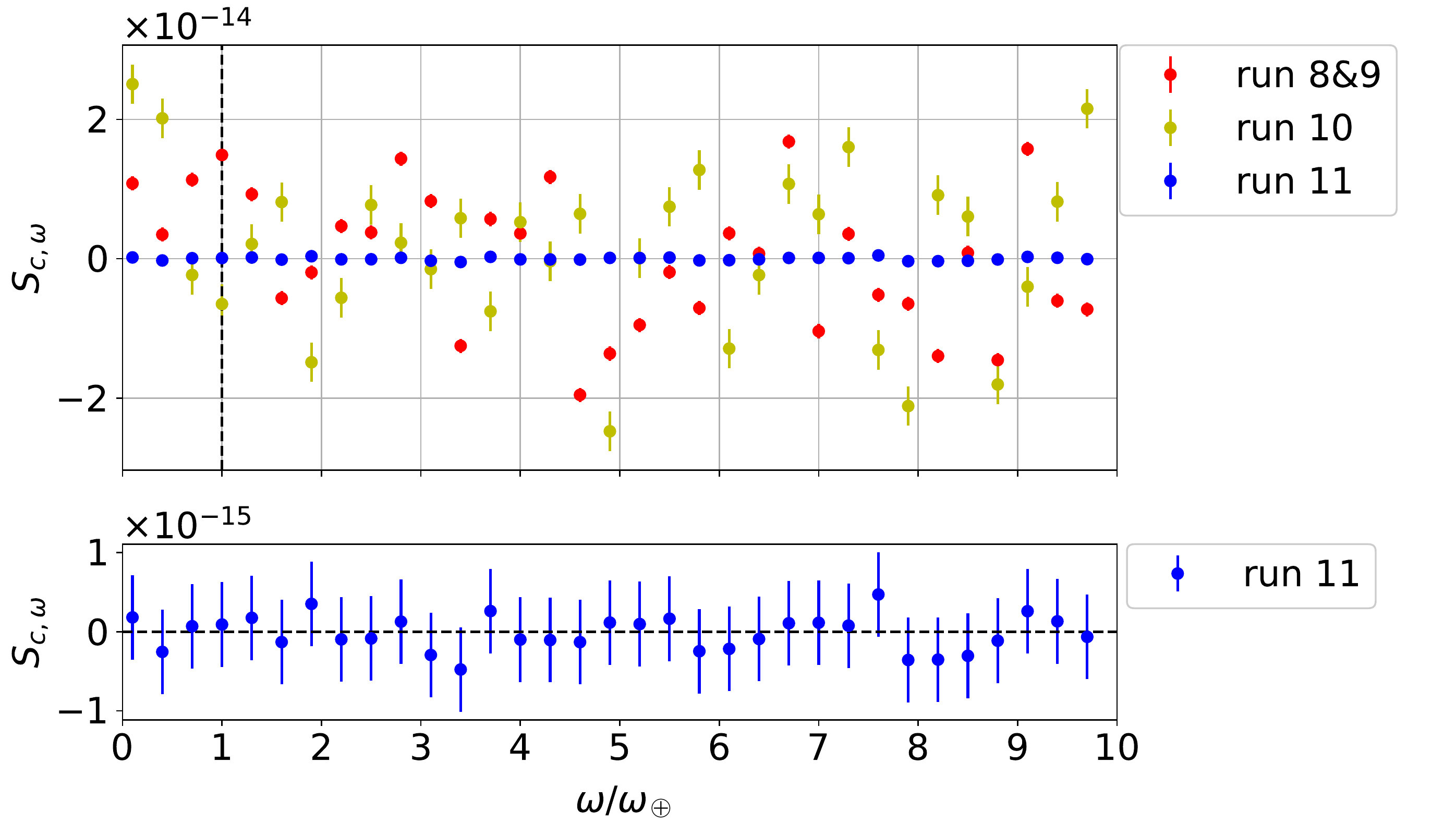}
\caption{The fitted $S_{c,\omega}$ coefficient with $N_r = 10$ as a function of normalized frequency $\omega / \omega_\oplus$ for the different experimental runs.}
\label{compareRuns}
\end{figure}

To get an indication of the sensitivity of our experiment, values for $S_{c,\omega}$ were fitted from the data (as in figure \ref{SandCvsDay} for run 11) and are shown in Figure \ref{compareRuns} as a function of normalised frequency (with respect to sidereal) for all experimental runs. It is clear that runs 8 to 10 are limited by an extra noise process that scatters the excursions from zero at an amplitude greater than the standard errors of the fitting, while in run 11 the systematic was eliminated.

To understand systematics we undertook measurements of the most likely parameters that would influence the measurements. During run 10 the temperature was continuously monitored in unison with the experiment. Data files under went the same process as the demodulated least squares. Results revealed no significant temperature effects at the rotation or sidereal frequencies above the standard error of fluctuations, so this effect was ruled out. The amplitude of the rotation systematics proved to be sensitive to magnetic field, measurements were made to measure the magnetic field in the laboratory as a function of time. However, it was shown that the magnetic field was not the cause of the extra systematic fluctuation as observed in Figure \ref{compareRuns} for runs 8 to 10.

The limitation of runs 8 to 10 was shown to come from the fact that the data acquisition runs on a non-realtime operating system relying on WiFi data acquisition. The operating system limits sampling stability and, as we learned, may result in smearing of the systematic signals that can eventually limit the performance. Furthermore, the wireless network running on month time scales may interrupt the data acquisition randomly leaving substantial (a few seconds) data gaps. Although, these interruptions on their own do not directly limit the system estimated performance, they do not allow application of direct spectral methods as applied in our original experiment\cite{Lo:2016aa}, as these methods assume uniformity of the time scale and thus cannot be directly applied to data with gaps (unlike the implementation of the least squares method). 

To reduce these systematic effects related to the stability of the timing on the WiFi based acquisition channel, the amount of data retrieved from the digitizer at each acquisition step was reduced for run 11 by 40\%, which in principle could deteriorate the signal to noise ratio due to decimation of data. However, due to the high frequency of the acquisition (1.6 kHz sampling rate) this deterioration only happens primarily at higher frequencies when compared to the rotation frequency where the data is dominated by the measurement apparatus noise floor. Thus near the frequencies of interest this change has little effect. In the long run, the reduction in channel loadings demonstrated significant improvement in systematic effects related to the timing clearly observed in run 11.


We have also gone through the process of optimizing $N_r$ at $2\omega_R$ for the first stage of the DLS technique to end up with the best signal to noise ratio at the end of the entire process. The optimized value of $N_r$ minimizes the standard error of the fitted SME coefficients, and can be interpreted as a balance between two processes, the broad band white noise in the system, and the narrow band noise due to fluctuations of the stability of the rotation (systematic noise). When the system is dominated by white noise, averaging by fitting over a large number of rotations helps to reduce the standard error. However, if the narrowband rotation systematic fluctuates the fitting to the systematic at the rotation harmonic becomes more uncertain at large values of $N_r$. The optimal value is attained when the noise contributed by both processes is equal \cite{Stanwix2006}. While the white noise is nearly constant throughout all relevant Fourier frequencies, the systemic noise can vary from run to run. As we can see from Figure \ref{compareRuns}, our experiment is mostly limited by systemic noise in runs 8 to 10 with $N_r=10$ over all fitted frequencies. The relevant frequencies to search for LIV are at the values $\omega / \omega_\oplus$ =1 or 2. In general runs 8 to 10 show significant results with respect to the precision of the experiment (standard error) however, if this was due to LIV and not an added systematic noise process one would expect no significance at all other frequencies, which is clearly not the case. By increasing $N_r$ to 1000 the precision of the experiment is reduced by an order of magnitude, but the averaging time is large enough to cause the effect of the systematic fluctuations to be effectively random (not significant). In contrast, the improvements made for run 11 give an optimum value of $N_r=1000$, so the rotation systematic has become stable enough that the results are no longer significant and a much better precision is achieved. We have determined that the amplitude of the rotation systematics is limited by magnetic field sensitivity. In the next runs shielding will be provided to reduce this effect.

\section{Further Work and Perspectives}

Based on the results of run 11, we estimate that LIV SME matter coefficients could have limits set of order $10^{-16}$ GeV with a years worth of data, due to the improvements detailed in this work. This experimental run will be finished during the second half of 2018.  Due to the direct dependence of the experiment on the neutron sector, this experiment has an advantage over recent atomic clock experiments\cite{guena2010} that tests matter sector LIV coefficients\cite{WolfPRL2006,WolfPRD2017}. Despite the better sensitivity, the atomic clock experiments are mainly sensitive to the proton, with suppressed sensitivity to neutrons by a factor of 0.021, and thus can only put limits on a linear combination of the two particles \cite{WolfPRD2017}. In the same paper it is shown by using the Schmidt model, proton coefficients can have independent limits set at a sensitivity better than the limits set by the phonon sector experiments. Thus, the phonon sector experiment may ignore the proton sector and put limits only in the neutron sector, as was achieved in the prior experiment \cite{Lo:2016aa}.

Further improvement of the experimental sensitivity by improving the frequency stability of quartz oscillators at room temperature is doubtful, because the technology has reached its limits over past couple of decades. Frequency stability of 5 and 10 MHz quartz oscillators is limited by the intrinsic flicker noise of BAW resonators. An alternative solution would be a transition from room temperature operation to liquid helium environment where BAW quartz resonators demonstrate orders of magnitude improvement in quality factors\cite{Lo:2016aa,quartzPRL,ScRep}, which in principle could lead to a three orders of magnitude improvement. Although development of frequency standards based on cryogenic quartz resonators is associated with several technological challenges\cite{Goryachev:2012jx,Goryachev:2013ly}.

The data from this experiment can be easily adapted to search for higher dimensions SME LIV coefficients in the phonon/matter sector in a similar way that has been implemented for rotating sapphire oscillators in the photon sector \cite{Parker2011,Mewes2012,Michimura2013,Parker2015}. This will most likely to include the analysis of a range of other harmonics of the rotation and sidereal frequencies, as has been done in the past in the photon sector. 

\section*{Acknowledgements}

This work was supported by the Australian Research Council grant number CE170100009 and DP160100253 as well as the Austrian Science Fund (FWF) J3680. We thank Paul Stanwix for some help with the data analysis.

\ifCLASSOPTIONcaptionsoff
  \newpage
\fi

\bibliographystyle{IEEEtran}
%\bibliography{biblio}

\begin{thebibliography}{10}
\providecommand{\url}[1]{#1}
\csname url@samestyle\endcsname
\providecommand{\newblock}{\relax}
\providecommand{\bibinfo}[2]{#2}
\providecommand{\BIBentrySTDinterwordspacing}{\spaceskip=0pt\relax}
\providecommand{\BIBentryALTinterwordstretchfactor}{4}
\providecommand{\BIBentryALTinterwordspacing}{\spaceskip=\fontdimen2\font plus
\BIBentryALTinterwordstretchfactor\fontdimen3\font minus
  \fontdimen4\font\relax}
\providecommand{\BIBforeignlanguage}[2]{{%
\expandafter\ifx\csname l@#1\endcsname\relax
\typeout{** WARNING: IEEEtran.bst: No hyphenation pattern has been}%
\typeout{** loaded for the language `#1'. Using the pattern for}%
\typeout{** the default language instead.}%
\else
\language=\csname l@#1\endcsname
\fi
#2}}
\providecommand{\BIBdecl}{\relax}
\BIBdecl

\bibitem{ColladayKostelecky}
\BIBentryALTinterwordspacing
D.~Colladay and V.~A. Kosteleck{\'y}, ``Cpt violation and the standard model,''
  \emph{Phys. Rev. D}, vol.~55, no.~11, pp. 6760--6774, 06 1997. [Online].
  Available: \url{http://link.aps.org/doi/10.1103/PhysRevD.55.6760}
\BIBentrySTDinterwordspacing

\bibitem{KosteleckyLane}
\BIBentryALTinterwordspacing
V.~A. Kosteleck\'y and C.~D. Lane, ``Constraints on lorentz violation from
  clock-comparison experiments,'' \emph{Phys. Rev. D}, vol.~60, p. 116010, Nov
  1999. [Online]. Available:
  \url{http://link.aps.org/doi/10.1103/PhysRevD.60.116010}
\BIBentrySTDinterwordspacing

\bibitem{MM2}
A.~Michelson and E.~Morley, ``On the relative motion of the earth and the
  luminiferous {\ae}ther,'' \emph{Phil. Mag.}, vol.~24, p. 449, 1887.

\bibitem{MM3}
R.~Kennedy and E.~Thorndike, ``Experimental establishment of the relativity of
  time,'' \emph{Phys. Rev.}, vol.~42, p. 400, 1932.

\bibitem{KosteleckyMewesPRD}
\BIBentryALTinterwordspacing
V.~A. Kosteleck{\'y} and M.~Mewes, ``Signals for lorentz violation in
  electrodynamics,'' \emph{Phys. Rev. D}, vol.~66, no.~5, pp. 056\,005--, 09
  2002. [Online]. Available:
  \url{http://link.aps.org/doi/10.1103/PhysRevD.66.056005}
\BIBentrySTDinterwordspacing

\bibitem{Stanwix}
\BIBentryALTinterwordspacing
P.~L. Stanwix, M.~E. Tobar, P.~Wolf, M.~Susli, C.~R. Locke, E.~N. Ivanov,
  J.~Winterflood, and F.~van Kann, ``Test of lorentz invariance in
  electrodynamics using rotating cryogenic sapphire microwave oscillators,''
  \emph{Phys. Rev. Lett}, vol.~95, no.~4, p. 040404, 07 2005. [Online].
  Available: \url{http://link.aps.org/doi/10.1103/PhysRevLett.95.040404}
\BIBentrySTDinterwordspacing

\bibitem{Hohensee}
\BIBentryALTinterwordspacing
M.~A. Hohensee, P.~L. Stanwix, M.~E. Tobar, S.~R. Parker, D.~F. Phillips, and
  R.~L. Walsworth, ``Improved constraints on isotropic shift and anisotropies
  of the speed of light using rotating cryogenic sapphire oscillators,''
  \emph{Phys. Rev. D}, vol.~82, no.~7, pp. 076\,001--, 10 2010. [Online].
  Available: \url{http://link.aps.org/doi/10.1103/PhysRevD.82.076001}
\BIBentrySTDinterwordspacing

\bibitem{Schiller}
\BIBentryALTinterwordspacing
C.~Eisele, A.~Y. Nevsky, and S.~Schiller, ``Laboratory test of the isotropy of
  light propagation at the {\$}{\{}10{\}}\^{}{\{}-17{\}}{\$} level,''
  \emph{Phys. Rev. Lett}, vol. 103, no.~9, pp. 090\,401--, 08 2009. [Online].
  Available: \url{http://link.aps.org/doi/10.1103/PhysRevLett.103.090401}
\BIBentrySTDinterwordspacing

\bibitem{Achim}
\BIBentryALTinterwordspacing
S.~Herrmann, A.~Senger, K.~M{\"o}hle, M.~Nagel, E.~V. Kovalchuk, and A.~Peters,
  ``Rotating optical cavity experiment testing lorentz invariance at the
  {\$}{\{}10{\}}\^{}{\{}-17{\}}{\$} level,'' \emph{Phys. Rev. D}, vol.~80,
  no.~10, pp. 105\,011--, 11 2009. [Online]. Available:
  \url{http://link.aps.org/doi/10.1103/PhysRevD.80.105011}
\BIBentrySTDinterwordspacing

\bibitem{Haeffner}
\BIBentryALTinterwordspacing
T.~Pruttivarasin, M.~Ramm, S.~G. Porsev, I.~I. Tupitsyn, M.~S. Safronova, M.~A.
  Hohensee, and H.~Haffner, ``Michelson-morley analogue for electrons using
  trapped ions to test lorentz symmetry,'' \emph{Nature}, vol. 517, no. 7536,
  pp. 592--595, 01 2015. [Online]. Available:
  \url{http://dx.doi.org/10.1038/nature14091}
\BIBentrySTDinterwordspacing

\bibitem{Nagel}
\BIBentryALTinterwordspacing
M.~Nagel, S.~R. Parker, E.~V. Kovalchuk, P.~L. Stanwix, J.~G. Hartnett, E.~N.
  Ivanov, A.~Peters, and M.~E. Tobar, ``Direct terrestrial test of lorentz
  symmetry in electrodynamics to 10-18,'' \emph{Nat Commun}, vol.~6, 09 2015.
  [Online]. Available: \url{http://dx.doi.org/10.1038/ncomms9174}
\BIBentrySTDinterwordspacing

\bibitem{Coleman1999}
\BIBentryALTinterwordspacing
S.~Coleman and S.~L. Glashow, ``High-energy tests of lorentz invariance,''
  \emph{Phys. Rev. D}, vol.~59, no.~11, pp. 116\,008--, 04 1999. [Online].
  Available: \url{http://link.aps.org/doi/10.1103/PhysRevD.59.116008}
\BIBentrySTDinterwordspacing

\bibitem{Gomes}
\BIBentryALTinterwordspacing
A.~Gomes, V.~A. Kosteleck{\'y}, and A.~J. Vargas, ``Laboratory tests of lorentz
  and {\$}cpt{\$} symmetry with muons,'' \emph{Phys. Rev. D}, vol.~90, no.~7,
  pp. 076\,009--, 10 2014. [Online]. Available:
  \url{http://link.aps.org/doi/10.1103/PhysRevD.90.076009}
\BIBentrySTDinterwordspacing

\bibitem{Bear}
\BIBentryALTinterwordspacing
D.~Bear, R.~E. Stoner, R.~L. Walsworth, V.~A. Kosteleck{\'y}, and C.~D. Lane,
  ``Limit on lorentz and {$\backslash$}textit{\{}cpt{\}} violation of the
  neutron using a two-species noble-gas maser,'' \emph{Phys. Rev. Lett},
  vol.~85, no.~24, pp. 5038--5041, 12 2000. [Online]. Available:
  \url{http://link.aps.org/doi/10.1103/PhysRevLett.85.5038}
\BIBentrySTDinterwordspacing

\bibitem{Altschul}
\BIBentryALTinterwordspacing
B.~Altschul, ``Limits on neutron lorentz violation from the stability of
  primary cosmic ray protons,'' \emph{Phys. Rev. D}, vol.~78, no.~8, pp.
  085\,018--, 10 2008. [Online]. Available:
  \url{http://link.aps.org/doi/10.1103/PhysRevD.78.085018}
\BIBentrySTDinterwordspacing

\bibitem{KosteleckyTasson}
\BIBentryALTinterwordspacing
V.~A. Kosteleck{\'y} and J.~D. Tasson, ``Matter-gravity couplings and lorentz
  violation,'' \emph{Phys. Rev. D}, vol.~83, no.~1, pp. 016\,013--, 01 2011.
  [Online]. Available: \url{http://link.aps.org/doi/10.1103/PhysRevD.83.016013}
\BIBentrySTDinterwordspacing

\bibitem{Brown}
\BIBentryALTinterwordspacing
J.~M. Brown, S.~J. Smullin, T.~W. Kornack, and M.~V. Romalis, ``New limit on
  lorentz- and {\$}cpt{\$}-violating neutron spin interactions,'' \emph{Phys.
  Rev. Lett}, vol. 105, no.~15, pp. 151\,604--, 10 2010. [Online]. Available:
  \url{http://link.aps.org/doi/10.1103/PhysRevLett.105.151604}
\BIBentrySTDinterwordspacing

\bibitem{WolfPRL2006}
\BIBentryALTinterwordspacing
P.~Wolf, F.~Chapelet, S.~Bize, and A.~Clairon, ``Cold atom clock test of
  lorentz invariance in the matter sector,'' \emph{Phys. Rev. Lett}, vol.~96,
  no.~6, pp. 060\,801--, 02 2006. [Online]. Available:
  \url{http://link.aps.org/doi/10.1103/PhysRevLett.96.060801}
\BIBentrySTDinterwordspacing

\bibitem{datatables}
\BIBentryALTinterwordspacing
V.~A. Kosteleck{\'y} and N.~Russell, ``Data tables for lorentz and {\$}cpt{\$}
  violation,'' \emph{Reviews of Modern Physics, arXiv:0801.0287}, vol.~83,
  no.~1, pp. 11--31, 03 2011. [Online]. Available:
  \url{http://link.aps.org/doi/10.1103/RevModPhys.83.11}
\BIBentrySTDinterwordspacing

\bibitem{Lo:2016aa}
\BIBentryALTinterwordspacing
A.~Lo, P.~Haslinger, E.~Mizrachi, L.~Anderegg, H.~M{\"u}ller, M.~Hohensee,
  M.~Goryachev, and M.~E. Tobar, ``Acoustic tests of lorentz symmetry using
  quartz oscillators,'' \emph{Physical Review X}, vol.~6, no.~1, pp.
  011\,018--, 02 2016. [Online]. Available:
  \url{https://link.aps.org/doi/10.1103/PhysRevX.6.011018}
\BIBentrySTDinterwordspacing

\bibitem{QCM1}
D.~Johannsmann, \emph{The Quartz Crystal Microbalance in Soft Matter
  Research}.\hskip 1em plus 0.5em minus 0.4em\relax Switzerland: Springer
  International Publishing, 2015.

\bibitem{Salzenstein:2010aa}
P.~Salzenstein, A.~Kuna, L.~Sojdr, and J.~Chauvin, ``Significant step in
  ultra-high stability quartz crystal oscillators,'' \emph{Electron. Lett.},
  vol.~46, no.~21, pp. 1433--1434, October 2010.

\bibitem{Vig:1991aa}
J.~R. Vig and T.~R. Meeker, ``The aging of bulk acoustic wave resonators,
  filters and oscillators,'' \emph{Proceedings of the 45th Annual Symposium on
  Frequency Control 1991}, pp. 77--101, 1991.

\bibitem{Filler:1988oa}
R.~L. Filler, ``The acceleration sensitivity of quartz crystal oscillators: a
  review,'' \emph{Ultrasonics, Ferroelectrics and Frequency Control, IEEE
  Transactions on}, vol.~35, no.~3, pp. 297--305, May 1988.

\bibitem{Besson:1979aa}
R.~Besson, J.~J. Gagnepain, D.~Janiaud, and M.~Valdois, ``Design of a bulk wave
  resonator insensitive to acceleration,'' \emph{33rd Annual Symposium on
  Frequency Control}, pp. 337--345, 1979.

\bibitem{Ivanov:1998aa}
E.~N. Ivanov, M.~E. Tobar, and R.~A. Woode, ``Microwave interferometry:
  application to precision measurements and noise reduction techniques,''
  \emph{IEEE Transactions on Ultrasonics, Ferroelectrics, and Frequency
  Control}, vol.~45, no.~6, pp. 1526--1536, 1998.

\bibitem{Wolf2003}
\BIBentryALTinterwordspacing
P.~Wolf, S.~Bize, A.~Clairon, A.~N. Luiten, G.~Santarelli, and M.~E. Tobar,
  ``Tests of lorentz invariance using a microwave resonator,'' \emph{Phys. Rev.
  Lett.}, vol.~90, p. 060402, Feb 2003. [Online]. Available:
  \url{https://link.aps.org/doi/10.1103/PhysRevLett.90.060402}
\BIBentrySTDinterwordspacing

\bibitem{Wolf2004}
\BIBentryALTinterwordspacing
P.~Wolf, S.~Bize, A.~Clairon, G.~Santarelli, M.~E. Tobar, and A.~N. Luiten,
  ``Improved test of lorentz invariance in electrodynamics,'' \emph{Phys. Rev.
  D}, vol.~70, p. 051902, Sep 2004. [Online]. Available:
  \url{https://link.aps.org/doi/10.1103/PhysRevD.70.051902}
\BIBentrySTDinterwordspacing

\bibitem{Tobar2010}
\BIBentryALTinterwordspacing
M.~E. Tobar, P.~Wolf, S.~Bize, G.~Santarelli, and V.~Flambaum, ``Testing local
  lorentz and position invariance and variation of fundamental constants by
  searching the derivative of the comparison frequency between a cryogenic
  sapphire oscillator and hydrogen maser,'' \emph{Phys. Rev. D}, vol.~81, p.
  022003, Jan 2010. [Online]. Available:
  \url{https://link.aps.org/doi/10.1103/PhysRevD.81.022003}
\BIBentrySTDinterwordspacing

\bibitem{Tobar2013}
\BIBentryALTinterwordspacing
M.~E. Tobar, P.~L. Stanwix, J.~J. McFerran, J.~Gu\'ena, M.~Abgrall, S.~Bize,
  A.~Clairon, P.~Laurent, P.~Rosenbusch, D.~Rovera, and G.~Santarelli,
  ``Testing local position and fundamental constant invariance due to periodic
  gravitational and boost using long-term comparison of the syrte atomic
  fountains and h-masers,'' \emph{Phys. Rev. D}, vol.~87, p. 122004, Jun 2013.
  [Online]. Available:
  \url{https://link.aps.org/doi/10.1103/PhysRevD.87.122004}
\BIBentrySTDinterwordspacing

\bibitem{Abgrall2016}
M.~Abgrall, J.~Gu{\'e}na, M.~Lours, G.~Santarelli, M.~E. Tobar, S.~Bize,
  S.~Grop, B.~Dubois, C.~Fluhr, and V.~Giordano, ``High-stability comparison of
  atomic fountains using two different cryogenic oscillators,'' \emph{IEEE
  Transactions on Ultrasonics, Ferroelectrics, and Frequency Control}, vol.~63,
  no.~8, pp. 1198--1203, Aug 2016.

\bibitem{Stanwix2006}
\BIBentryALTinterwordspacing
P.~L. Stanwix, M.~E. Tobar, P.~Wolf, C.~R. Locke, and E.~N. Ivanov, ``Improved
  test of lorentz invariance in electrodynamics using rotating cryogenic
  sapphire oscillators,'' \emph{Phys. Rev. D}, vol.~74, p. 081101, Oct 2006.
  [Online]. Available:
  \url{https://link.aps.org/doi/10.1103/PhysRevD.74.081101}
\BIBentrySTDinterwordspacing

\bibitem{Meuller2007}
\BIBentryALTinterwordspacing
H.~M\"uller, P.~L. Stanwix, M.~E. Tobar, E.~Ivanov, P.~Wolf, S.~Herrmann,
  A.~Senger, E.~Kovalchuk, and A.~Peters, ``Tests of relativity by
  complementary rotating michelson-morley experiments,'' \emph{Phys. Rev.
  Lett.}, vol.~99, p. 050401, Jul 2007. [Online]. Available:
  \url{https://link.aps.org/doi/10.1103/PhysRevLett.99.050401}
\BIBentrySTDinterwordspacing

\bibitem{Tobar2009}
\BIBentryALTinterwordspacing
M.~E. Tobar, E.~N. Ivanov, P.~L. Stanwix, J.-M.~G. le~Floch, and J.~G.
  Hartnett, ``Rotating odd-parity lorentz invariance test in electrodynamics,''
  \emph{Phys. Rev. D}, vol.~80, p. 125024, Dec 2009. [Online]. Available:
  \url{https://link.aps.org/doi/10.1103/PhysRevD.80.125024}
\BIBentrySTDinterwordspacing

\bibitem{Parker2011}
\BIBentryALTinterwordspacing
S.~R. Parker, M.~Mewes, P.~L. Stanwix, and M.~E. Tobar, ``Cavity bounds on
  higher-order lorentz-violating coefficients,'' \emph{Phys. Rev. Lett.}, vol.
  106, p. 180401, May 2011. [Online]. Available:
  \url{https://link.aps.org/doi/10.1103/PhysRevLett.106.180401}
\BIBentrySTDinterwordspacing

\bibitem{guena2010}
J.~Guena, P.~Rosenbusch, P.~Laurent, M.~Abgrall, D.~Rovera, G.~Santarelli,
  M.~E. Tobar, S.~Bize, and A.~Clairon, ``Demonstration of a dual alkali rb/cs
  fountain clock,'' \emph{IEEE Transactions on Ultrasonics, Ferroelectrics, and
  Frequency Control}, vol.~57, no.~3, pp. 647--653, March 2010.

\bibitem{WolfPRD2017}
\BIBentryALTinterwordspacing
H.~Pihan-Le~Bars, C.~Guerlin, R.-D. Lasseri, J.-P. Ebran, Q.~G. Bailey,
  S.~Bize, E.~Khan, and P.~Wolf, ``Lorentz-symmetry test at planck-scale
  suppression with nucleons in a spin-polarized $^{133}\mathrm{Cs}$ cold atom
  clock,'' \emph{Phys. Rev. D}, vol.~95, p. 075026, Apr 2017. [Online].
  Available: \url{https://link.aps.org/doi/10.1103/PhysRevD.95.075026}
\BIBentrySTDinterwordspacing

\bibitem{quartzPRL}
M.~Goryachev, D.~Creedon, S.~Galliou, and M.~Tobar, ``Observation of rayleigh
  phonon scattering through excitation of extremely high overtones in low-loss
  cryogenic acoustic cavities for hybrid quantum systems,'' \emph{Phys. Rev.
  Lett}, vol. 111, no.~8, p. 085502, 2013.

\bibitem{ScRep}
S.~Galliou, M.~Goryachev, R.~Bourquin, P.~Abbe, J.~Aubry, and M.~Tobar,
  ``Extremely low loss phonon-trapping cryogenic acoustic cavities for future
  physical experiments,'' \emph{Nature: Scientific Reports}, vol.~3, no. 2132,
  2013.

\bibitem{Goryachev:2012jx}
M.~Goryachev, S.~Galliou, P.~Abbe, P.~Bourgeois, S.~Grop, and B.~Dubois,
  ``Quartz resonator instabilities under cryogenic conditions,''
  \emph{Ultrasonics, Ferroelectrics and Frequency Control, IEEE Transactions
  on}, vol.~59, no.~1, pp. 21--29, January 2012.

\bibitem{Goryachev:2013ly}
\BIBentryALTinterwordspacing
M.~Goryachev, S.~Galliou, J.~Imbaud, and P.~Abb{\'e}, ``Advances in development
  of quartz crystal oscillators at liquid helium temperatures,''
  \emph{Cryogenics}, vol.~57, no.~0, pp. 104--112, 10 2013. [Online].
  Available:
  \url{http://www.sciencedirect.com/science/article/pii/S0011227513000532}
\BIBentrySTDinterwordspacing

\bibitem{Mewes2012}
\BIBentryALTinterwordspacing
M.~Mewes, ``Optical-cavity tests of higher-order lorentz violation,''
  \emph{Phys. Rev. D}, vol.~85, p. 116012, Jun 2012. [Online]. Available:
  \url{https://link.aps.org/doi/10.1103/PhysRevD.85.116012}
\BIBentrySTDinterwordspacing

\bibitem{Michimura2013}
\BIBentryALTinterwordspacing
Y.~Michimura, M.~Mewes, N.~Matsumoto, Y.~Aso, and M.~Ando, ``Optical cavity
  limits on higher order lorentz violation,'' \emph{Phys. Rev. D}, vol.~88, p.
  111101, Dec 2013. [Online]. Available:
  \url{https://link.aps.org/doi/10.1103/PhysRevD.88.111101}
\BIBentrySTDinterwordspacing

\bibitem{Parker2015}
\BIBentryALTinterwordspacing
S.~R. Parker, M.~Mewes, F.~N. Baynes, and M.~E. Tobar, ``Bounds on higher-order
  lorentz-violating photon sector coefficients from an asymmetric optical ring
  resonator experiment,'' \emph{Physics Letters A}, vol. 379, no.~42, pp. 2681
  -- 2684, 2015. [Online]. Available:
  \url{http://www.sciencedirect.com/science/article/pii/S0375960115006726}
\BIBentrySTDinterwordspacing

\end{thebibliography}

% Generated by IEEEtran.bst, version: 1.14 (2015/08/26)

\end{document}